\documentstyle[aps,preprint,amssymb,epsfig]{revtex}
\newcommand{\ga}{\alpha}
\newcommand{\gb}{\beta}
\newcommand{\gc}{\gamma}
\newcommand{\gd}{\delta}
\newcommand{\gk}{\kappa}

\newcommand{\gl}{\lambda}
\newcommand{\gs}{\sigma}

\tightenlines

\begin{document}
\title {The $\alpha$-particle in nuclear matter} \author{M. Beyer$^1$,
S.A.
  Sofianos$^2$, C. Kuhrts$^1$, G. R\"opke$^1$, and P. Schuck$^3$}
\address{$^1$Fachbereich Physik, Universit\"at Rostock, 18051 Rostock,
  Germany\\
  $^3$Dept. of Physics, University of South Africa, Pretoria 0003, South Africa\\
  $^3$Institut de Science Nucl\'eaires, Universit\'e Joseph Fourier,
  CNRS-IN2P3 53, Avenue des Martyrs, F-38026 Grenoble Cedex, France}
\maketitle
\begin{abstract}
  Among the light nuclear clusters
  the $\alpha$-particle is by far the strongest bound system and
  therefore expected to play a significant role in the dynamics of
  nuclei and the phases of nuclear matter. To systematically study the
  properties of the $\alpha$-particle we have derived an effective
   four-body equation of the   Alt-Grassberger-Sandhas (AGS) type 
   that includes the  dominant medium effects, i.e.  self energy 
  corrections and
  Pauli-blocking in a consistent way.  The equation is solved
  utilizing the energy dependent pole expansion for the subsystem amplitudes. 
   We find that the Mott
  transition of an  $\alpha$-particle at rest differs from that
  expected from perturbation theory and occurs at approximately 1/10
  of nuclear matter densities.
\end{abstract}

\section{Introduction}

The modification of few-body properties such as the binding energy and
the wave function of a bound state due to a medium of finite
temperature and density is an important subject of many-particle
theory. As an example we consider symmetric nuclear matter consisting
of nucleons (equal number of protons and neutrons) at density $\rho$ and
temperature $T$. The modification of single-nucleon properties can be
obtained from a Dyson equation in terms of a self-energy. In a
specific approximation the quasiparticle picture can be derived. A
more rigorous description leads to the nucleon spectral function.
Similarly we can consider the two-nucleon system where the medium
modification are obtained from a Bethe-Goldstone
equation~\cite{fet71}. In addition to the self-energy shift, also
Pauli-blocking has to be taken into account that is of the same order
of magnitude. It has been shown \cite{RMS,SRS} that, as a consequence,
deuterons in nuclear matter become unbound if the density exceeds a
certain value, the Mott density.

Of course, the same mechanisms are also responsible for the
modification of higher clusters embedded in nuclear matter. However,
the solution of the few-body in-medium equations where the effects of
the medium are accounted for by a density and temperature dependent
contribution to the Hamiltonian has only been done within perturbation
theory, see \cite{RMSS}. Therefore the results achieved for the energy
shifts and the Mott densities are only approximations.

Recently, rigorous methods have been used to find solutions for the
three-body problem in nuclear
matter~\cite{bey96,bey97,beyFB,schadow,kuhrts}. The Faddeev equations
are extended to include the effects of the medium, and the
corresponding Alt-Grassberger-Sandhas (AGS) equations have been
solved~\cite{AGS}.  Different properties of the three-nucleon system
in the medium such as the modification of the binding energy of the
three-nucleon bound state~\cite{schadow} and the medium modification
of the nucleon-deuteron break-up cross section~\cite{bey96,bey97} have
been calculated.

In the present letter we give first results of the solution of the
in-medium four-particle equation describing the modification of the
binding energy of the $\alpha$-particle in symmetric nuclear matter. An
AGS-type equation has been solved and the results will be compared with
those of perturbation theory.

Note that the four-particle correlations in low-density nuclear matter
are very important because of the large binding energy of the
$\alpha$-particle.  They have to be accounted for not only in
equilibrium
when considering the nuclear matter equation of state or the
contributions of correlations to the single-nucleon spectral function,
but also in nonequilibrium such as the light cluster formation in
heavy ion collisions.

\section{In-medium few-body equations}
The few-nucleon problem in nuclear matter can be treated using Green
function approaches. Within the cluster-mean field expansion
\cite{RMSS}, a self-consistent system of equations can be derived
describing a $n$-nucleon cluster moving in a mean field produced by
the equilibrium mixture of clusters with arbitrary nucleon number $m$.
A Dyson equation approach to describe clusters at finite temperatures
and densities has been given in~\cite{duk98}. However, the
self-consistent determination of the composition of the medium is a
very challenging task that is not solved until now. We will perform
the approximation where the correlations in the medium are neglected
so that the embedding nuclear matter is described by the equilibrium
distribution of quasiparticles (see also \cite{SRS} for the
two-particle problem, or \cite{bey96,bey97,beyFB,schadow} for the
three-particle problem).

The extension of this formalism to describe $n$-nucleon correlations
in nuclear matter will be given elsewhere. Here, we will give some
of the basic relations, which are direct generalizations of the
three-particle case.

Let the Hamiltonian of the system be given by
\begin{equation}
        H =\sum_{1} \frac{k_1^2}{2 m} a_1^\dagger a_{1}+\;\frac{1}{2}\;
        \sum_{12 1'2'}V_2(12,1'2')\;a^\dagger_1  a^\dagger_2  a_{2'}a_{1'}
\label{eqn:H}
\end{equation}
where $a_1$ etc. denotes the Heisenberg operator of the particle
that includes  quantum numbers such as spin $s_1$ and momentum  $k_1$.
The free resolvent $G_0$ for an $n$-particle cluster
is given in Matsubara-Fourier representation by
\begin{equation}
        G_0(z) = (z-H_0)^{-1}\;N \equiv R_0(z)\;N,
\label{eqn:G0}
\end{equation}
where $G_0$, $H_0$, and $N$ is  a compact notation for matrices in the
space of $n$ particles with respect to the particle indices given
below.  Here $z$ denotes the Matsubara frequencies
$z_\gl=\pi\gl/(-i\gb)+\mu$ with
$\gl=0,\pm 2,\pm 4,\dots$ for bosons and $\gl=\pm 1,\pm 3,\dots$ in
the case of fermions. 
To simplify the notation we have further
dropped the index $n$ on the matrices, however use it if explicitly
needed.  The effective in-medium Hamiltonian $H_0$ for noninteracting
quasi-particles is given by
\begin{equation}
        H_0 = \sum_{i=1}^n
        \frac{k_i^2}{2m} +\Sigma_i\equiv \sum_{i=1}^n \varepsilon_i
\end{equation}
where the energy shift $\Sigma_1$ and the Fermi function $f_1$ are
\begin{eqnarray}
        \Sigma_1&=&\sum_{2}V_2(12,\widetilde{12})f_2,\\
        f_1&\equiv& f(\varepsilon_1) = \frac{1}{\mbox{e}^{(\varepsilon_1 -
        \mu)/k_BT}+1}.
\end{eqnarray}
The notation $\widetilde{12}$ means antisymmetrisation.
The factor $N$ in Eq.~(\ref{eqn:G0}) resembles the Pauli blocking or
normalization of the Green functions. This factor is different for the
different clusters considered depending on the number of particles $n$. 
It is given by
\begin{equation}
        N=\bar f_1\bar f_2 \dots \bar f_n\pm f_1f_2\dots f_n
\end{equation}
where  $\bar f=1-f$. The upper sign is for an odd number of fermions (Fermi type) 
and the lower for an even number of fermions (Bose type). Note that
$NR_0=R_0N$.

The full resolvent after
Matsubara-Fourier transformation may be written in the following way
\begin{equation}
        G(z)=(z-H_0-V)^{-1}N\equiv R(z)N,
\end{equation}
where the potential $V$ is a sum of two-body interactions
between pairs $\ga$, i.e.
\begin{equation}
        V=\sum_\ga V_\ga= \sum_\ga N_2^{\ga}V_2^{\ga},
\label{eqn:V2}
\end{equation}
and $V_2^{\ga}$ is the two-body potential given in Eq.~(\ref{eqn:H}).
The sum runs over all unique pairs in the cluster. Note, that as a
consequence of Eq.~(\ref{eqn:V2}) $V^\dagger\neq V$, also $R(z)N\neq
NR(z)$ that later on leads to right and left eigen--vectors.

To be more specific: If the interaction is between particle 1 and 2
(in the pair $\ga=(12)$ of a cluster of $n$ particles) the effective
potential of Eq.~(\ref{eqn:V2}) reads
\begin{equation}
\label{pauli}
        \langle 12|N_2^{(12)}V_2^{(12)}|1'2'\rangle = (\bar f_1\bar f_2 -
        f_1f_2)V_2(12,1'2').
\end{equation}
A useful notion is the channel resolvent $G_\ga(z)$ for an $n$
particle cluster, where only the pair interaction in channel $\ga$ is
considered.  This may be written as
\begin{eqnarray}
        G_\ga(z)&=&(z-H_0-V_\ga)^{-1}N\\
        &=&(z-H_0-N_2^\ga V_2^\ga)^{-1}N\equiv R_\ga(z)N.
\nonumber
\end{eqnarray}
Using $R^{-1}_0(z)$, $R^{-1}_\ga(z)$, and $R^{-1}(z)$ it is possible
to formally derive the resolvent equations in the standard way.  To
keep the formal equivalence to the isolated case,  the $n$-particle
channel $t$-matrix $T_\ga$ is defined by
\begin{equation}\label{eqn:defRa}
        R_\ga(z)=R_0(z)+R_0(z)T_\ga(z) R_0(z)\,.
\end{equation}
With the use of $T_\ga(z)=N_2^\ga T_2^\ga(z)$ Eq.~(\ref{eqn:defRa}) 
leads to the well-known Bethe-Goldstone equation~\cite{fet71}
\begin{equation}
        T_2^\ga(z)=V_2^\ga + V_2^\ga R_0(z) N_2^\ga  T_2^\ga (z)
        =V_2^\ga +  T_2^\ga (z) R_0(z) N_2^\ga V_2^\ga.
\label{eqn:T2}
\end{equation}
We remark that similiar equations have been written down by various authors
previously~\cite{SchuckR,SchuckR1}.

Note that the above equations are also valid for the two-particle
subsystem embedded in a larger cluster (three, four, or more
particles). As for the isolated equations the effects of the other
particles appear only in the Matsubara frequencies $z$ (energies) of
the other particles in the cluster. No additional blocking factors $N$
related to the larger cluster arise. Also note, that the changes due
to the Pauli blocking are in the resolvent $G_0$ not in the potential
$V_2$. However, it is possible to rewrite this equation and introduce
an effective potential as seen in Eq.~(\ref{eqn:T2}) and use
unchanged resolvents instead. 
Making use of the more intuitive picture
of a blocking in the propagation of the particles (related to the
resolvents) we find by
Eq.~(\ref{eqn:T2}) the correct expression for the $t$-matrix that
enters into the Boltzmann collision integral 
(see, for example, ~\cite{roepke}).

The derivation of the three-body equation is straight forward and has
been given elsewhere~\cite{bey97,beyFB,schadow,kuhrts}. The AGS operator
$U_{\gb\ga}(z)$~\cite{AGS} for the three
particle system is defined by
\begin{equation}\label{eqn:defUtr}
        R(z)=\gd_{\gb\ga}R_\ga(z) + R_\gb(z) U_{\gb\ga}(z) R_\ga(z).
\end{equation}
Inserting Eqs.~(\ref{eqn:defRa}) and~(\ref{eqn:T2}) in the above
identity we result with the AGS-type equation 
\begin{equation}
        U_{\gb\ga}(z)=\bar\gd_{\gb\ga}R_0(z)^{-1}+\sum_\gc\bar\gd_{\gb\gc}
        N_2^\gc T_2^\gc(z) R_0(z) U_{\gc\ga}(z),
\label{eqn:AGS2}
\end{equation}
that includes now medium effects as Pauli blocking and self energy shifts.
We used the notation $\bar\gd_{\ga\gb}= 1-\gd_{\ga\gb}$.
This equation solves the three-body transition operator for a
three-particle cluster as well as for a three-particle cluster
embedded in a more-particle cluster, i.e. the effect of the other
particles in the cluster is again only in the Matsubara frequency
(energy) $z$. The definition of the transition operator given by
Eq.~(\ref{eqn:defUtr}) was chosen so
that no additional factor $N$ appears in the final equation. This
guarantees that the cluster equations are valid also if they are part
of a larger cluster. Thus, the two-body subsystem  $t$-matrix entering in
Eq.~(\ref{eqn:AGS2}) is the same as the one  given in Eq.~(\ref{eqn:T2}).
Therefore, it is possible to use all results of the few-body 'algebra',
in particular those based on cluster decomposition.

The in-medium bound state equation for an $n$-particle cluster follows
from the homogeneous Lippmann-Schwinger equation and is given by
\begin{equation}
        |\psi_B\rangle =R_0(E_B)V|\psi_B\rangle=
        R_0(E_B) \sum_{\gc} N_2^\gc V_2^\gc|\psi_B\rangle\,
\label{eqn:bound}
\end{equation}
where the sum is over all unique pairs in the cluster.
As shown in Ref.~\cite{schadow} for the three-body bound state it is
convenient to introduce form factors
\begin{equation}
        |F_\gb\rangle=\sum_{\gc=1}^3\bar\gd_{\gb\gc} N_2^\gc  V_2^\gc
|\psi_{B_3}\rangle
\end{equation}
that leads to the homogeneous in-medium AGS-type equation
\begin{equation}
        |F_\ga\rangle=\sum_{\gb=1}^3 \bar\gd_{\ga\gb} N_2^\gb 
        T_2^\gb R_0(B_3)|F_\gb\rangle.
\label{eqn:F3}
\end{equation}
We may generalize the  AGS method given in Refs.~\cite{san74,alt72,san75} 
to the in-medium  four-body case~\cite{sofianos} 
\begin{equation}
        |\psi_\gb\rangle = R_0(B_4) N_2^\gb T_2^\gb(B_4) R_0(B_4) \sum_{\gc=1}^6
        \bar\gd_{\gb\gc} R_0^{-1}(B_4)|\psi_\gc\rangle,
        \quad\gb=1,\dots, 6
\label{eqn:four}
\end{equation}
where
\begin{equation}
        |\psi_\gb\rangle=R_0(B_4) N_2^\gb V_2^\gb |\psi_{B_4}\rangle.
\end{equation}
Introducing the $3+1$ and $2+2$ cluster decomposition of the four-body system,
denoted by $\tau,\gs,\dots$,   the sum on the
right hand side of Eq.~(\ref{eqn:four}) may be rearranged by introducing
four-body form factors
\begin{equation}
        |{\cal F}_\gb^\gs\rangle = \sum_\tau \bar\gd_{\gs\tau}
        \sum_\ga \bar\gd^\tau_{\gb\ga} R_0^{-1}(B_4) |\psi_\ga\rangle
\end{equation}
with $\gb\subset\gs$, $\bar\gd^\tau_{\gb\ga}=\bar\gd_{\gb\ga}$, 
if $\gb,\ga\subset\tau$ and $\bar\gd^\tau_{\gb\ga}=0$ otherwise.
The homogeneous in-medium AGS-type equation for the four-body form
factors is then written    
\begin{equation}
        |{\cal F}^\gs_\gb\rangle=\sum_{\tau\gc} \bar\gd_{\gs\tau}
        U^\tau_{\gb\gc}(B_4)  R_0(B_4)  N_2^\gc T_2^\gc(B_4)
        R_0(B_4) |{\cal F}^\tau_\gc\rangle,
        \qquad\gb\subset\gs,\gc\subset\tau.
\label{eqn:hom}
\end{equation}
The driving kernel consists of the in-medium two-body $t$-matrix
defined  by the Bethe-Goldstone equation and the in-medium AGS-type
transition operator defined in Eq.~(\ref{eqn:AGS2}). Note also that an
additional Pauli blocking factor $N_2^\gc$ occurs.

The equations for the three-body scattering and bound state problem
have been solved numerically in
Refs.~\cite{bey96,bey97,schadow,kuhrts}. An exploratory
calculation to study a possible $\alpha$-like condensate (quartetting)
has been carried out using
a variational ansatz for the (2+2) channel  and by neglecting the (3+1)
channel~\cite{alpha}.

Because of the medium dependence of the equations the calculation time
increases drastically. This is due to the fact that  the positions of 
the deuteron pole as well as the three-nucleon pole vary with the intrinsic 
momentum and are not fixed at the usual binding energy because of the phase space
occupation through other particles.  Presently a sufficiently fast and
accurate method to solve the three- and four-body equations relies on
the separability of the subamplitutes that appear in the AGS
equations. To solve the four-body bound states we utilize the energy
dependent pole expansion (EDPE)~\cite{sofianos} that needs to be
adjusted to the in-medium case, because of different right and left
expansion functions  due to the nonsymmetric effective potential.

To be more specific,  we assume the following expansions for the amplitudes
of the respective sub-systems embedded in the four-body equation. For
the {\em two-body} sub-system we have
\begin{equation}
        T_\gc(z) \simeq \sum_n
        |\tilde\Gamma_{\gc n}(z)\rangle t_{\gc n}(z)\langle \Gamma_{\gc n}(z)|
        \simeq\sum_n |\tilde g_{\gc n}\rangle t_{\gc n}(z)\langle g_{\gc n}|
        =\sum_n
        N_2^\gc |g_{\gc n}\rangle t_{\gc n}(z)\langle g_{\gc n}|.
\label{eqn:Tsep2}
\end{equation}
The last equation of the right hand side is used in the present
calculation and reflects a simple Yamaguchi ansatz for the
form factors~\cite{yama}. The Pauli blocking factor then appears
explicitly. This has been used in Refs.~\cite{bey96,bey97,beyFB,kuhrts} and a
comparison to the Paris potential is given in Ref.\cite{schadow}.
For the present
purpose this approximation seems sufficient. For the {\em three-body}
subamplitudes we use  the EDPE  expansion
\begin{equation}
        \langle g_{\gb m}(z)| R_0(z) U^\tau_{\gb\gc}(z)R_0(z)| \tilde g_{\gc n}
        (z)\rangle
        \simeq \sum_{t,\mu\nu} |\tilde\Gamma^{\tau t, \mu}_{\gb m}(z)\rangle
        t^{\tau t}_{\mu\nu}(z)\langle \Gamma^{\tau t, \nu}_{\gc n}(z)|
\label{eqn:pole3}
\end{equation}
with
\begin{equation}
        |\tilde\Gamma^{\tau t, \mu}_{\gb m}(z)\rangle
        = \langle g_{\ga n}|R_0(z)| \tilde g_{\gb m}\rangle
        t_{\gb m}(B_3) |\tilde\Gamma^{\tau t, \mu}_{\gb m}\rangle\,.
\end{equation}
The Sturmian functions corresponding to the fixed energy $B_3$ are
given by
\begin{eqnarray}
        \eta_{t,\mu}|\tilde\Gamma^{\tau t, \mu}_{\ga n}\rangle&=&
        \sum_{\gb m}
        \langle g_{\ga n}|R_0(B_3)| \tilde g_{\gb m}\rangle
        t_{\gb m}(B_3)|\tilde\Gamma^{\tau t, \mu}_{\gb m}\rangle
\label{eqn:sturm}\\
        \eta_{t,\mu}|\Gamma^{\tau t, \mu}_{\ga n}\rangle&=&
        \sum_{\gb m}
        \langle \tilde g_{\ga n}|R_0(B_3)|  g_{\gb m}\rangle
        t_{\gb m}(B_3)|\Gamma^{\tau t, \mu}_{\gb m}\rangle\,.
\end{eqnarray}
Introducing the form factors
\begin{equation}
        |{\Bbb F}^{\gs s}_\mu\rangle
        = \sum_{\gb m}\langle\Gamma^{\gs s}_{\gb m,\nu}(B_4)|t_{\gb m}(B_4)
        \langle g_{\gb m}(B_4)| R_0(B_4)|{\cal F}^\gs_\gb\rangle
\end{equation}
%
we obtain the following homogeneous system of integral equations 
\begin{equation}
        |{\Bbb F}^{\gs s}_\mu\rangle =
        \sum_{\tau t}\sum_{\nu\gk} \sum_{\gc n} \bar\gd_{\gs\tau}
        \langle\Gamma^{\gs s, \nu}_{\gc n}(B_4)|t_{\gc n}(B_4)
        |\tilde\Gamma^{\tau t, \mu}_{\gc n}(B_4)\rangle\;
        t^{\tau t}_{\mu\gk}(B_4)\;|{\Bbb F}^{\tau t}_\gk\rangle.
\label{eqn:coup4}
\end{equation}
Formally these equations resemble the structure of the isolated
four-body equations. However, the dominant features of the influence
of the medium, i.e. the self-energy correction and the Pauli blocking,
are systematically taken into account.

Inclusion of spin-isospin degrees of freedom
and symmetrization is a challenging task for the
four-body problem and done as for the isolated case. To this
end we have intoduced angle averaged Pauli factors
as explained e.g. in Ref.~\cite{kuhrts} and fit the self-energy
by use of effective masses.

\section{Results and Conclusion}
To solve the four-body equation numerically we use a Yamaguchi type
rank one potential for the $^3S_1$ and $^1S_0$ channels. The
parameters are taken from an early work of Gibson and
Lehman~\cite{gibson}. We renormalized the calculated binding energy of the 
$\alpha$-particle so that it coincides for the 
isolated particle with the experimental one.
Presently, instead of using a more elaborated approach to the
isolated four-nucleon problem we merely study the {\em change} of the
binding energy due to the density and temperature of the surrounding
nuclear matter.  From our recent results for the three-body
systems~\cite{schadow}, we argue that the change is not very
sensitive to the particular form of the potential. We shall, therefore,
leave the study of model dependences for a future communication.

We calculated the binding energy of an $\alpha$-like cluster with zero
center of mass  momentum in symmetric nuclear matter at temperature $T$ = 10 MeV
as a function of the nucleon density. The results are shown as a solid
line in Fig.~1. The Mott transition occurs at a single
particle density of $\rho_{\rm Mott}=0.0207$ fm$^{-3}$. For comparison
we have given a perturbative calculation shown as dashed line. This
calculation is based on a simple Gaussian wave function for the
$\alpha$-particle with the width fitted to the electric rms radius.
Also the binding energy has been renormalized to the experimental
value. The Mott density in this case is at 0.0305 fm$^{-3}$ that
strongly differs from the value gained from the solution of
Eq.~(\ref{eqn:coup4}).

The corresponding curves for the triton and the deuteron are shown as
dotted and dashed dotted lines respectively. Note, that these binding
energies are for clusters at rest in the medium. Although it is an
interesting case, when the sub-clusters embedded in the larger cluster
vanishes as a bound state the sub-clusters in this case have a
dynamical binding energy that depends on the c.m. momentum.
Nevertheless, the question of Boromenian states and the Effimov effect
needs further investigation.

Unlike for the triton the $\alpha$-particle still exists at densities
where the Pauli blocking factor, see Eq.~(\ref{pauli}), becomes
negative. The usual procedure of symmetrizing the effective potential
by proper square root factors fails. Therefore when solving the
four-body equation we have to keep track of the right and left
eigenvectors in the subsystem.

In conclusion, we derived and  solved for the first time an effective in-medium
four-particle equation of the  AGS type. Applying it to symmetric nuclear 
matter, we found that the binding energy of the $\alpha$-particle decreases
with increasing
density due to Pauli blocking and disappears at a critical value of
the density (about 1/10 of the nuclear matter density for $T$ = 10
MeV). The dependence of the results on temperature and center of mass 
momentum will be the subject of an extended work.

\section*{Acknowledgement}
This work has been supported
by the Deutsche Forschungsgemeinschaft grant BE1092/7-1.


\newpage
\section*{Figure Captions}
\noindent
Fig. 1. Binding energy of an $\alpha$-like cluster with zero center of mass
      momentum embedded in symmetric nuclear matter at a temperature
      of $T=10$\,MeV as a function of nucleon density.  Solid line:
      Yamuguchi potential, renormalized to experimental binding energy
      at zero density.  Dashed line: perturbation approach. For
      comparison, the medium dependent binding energies of the
      deuteron (dashed-dotted) and triton (dotted) are also shown.

\begin{figure}[b]
  \begin{center}
    \epsfig{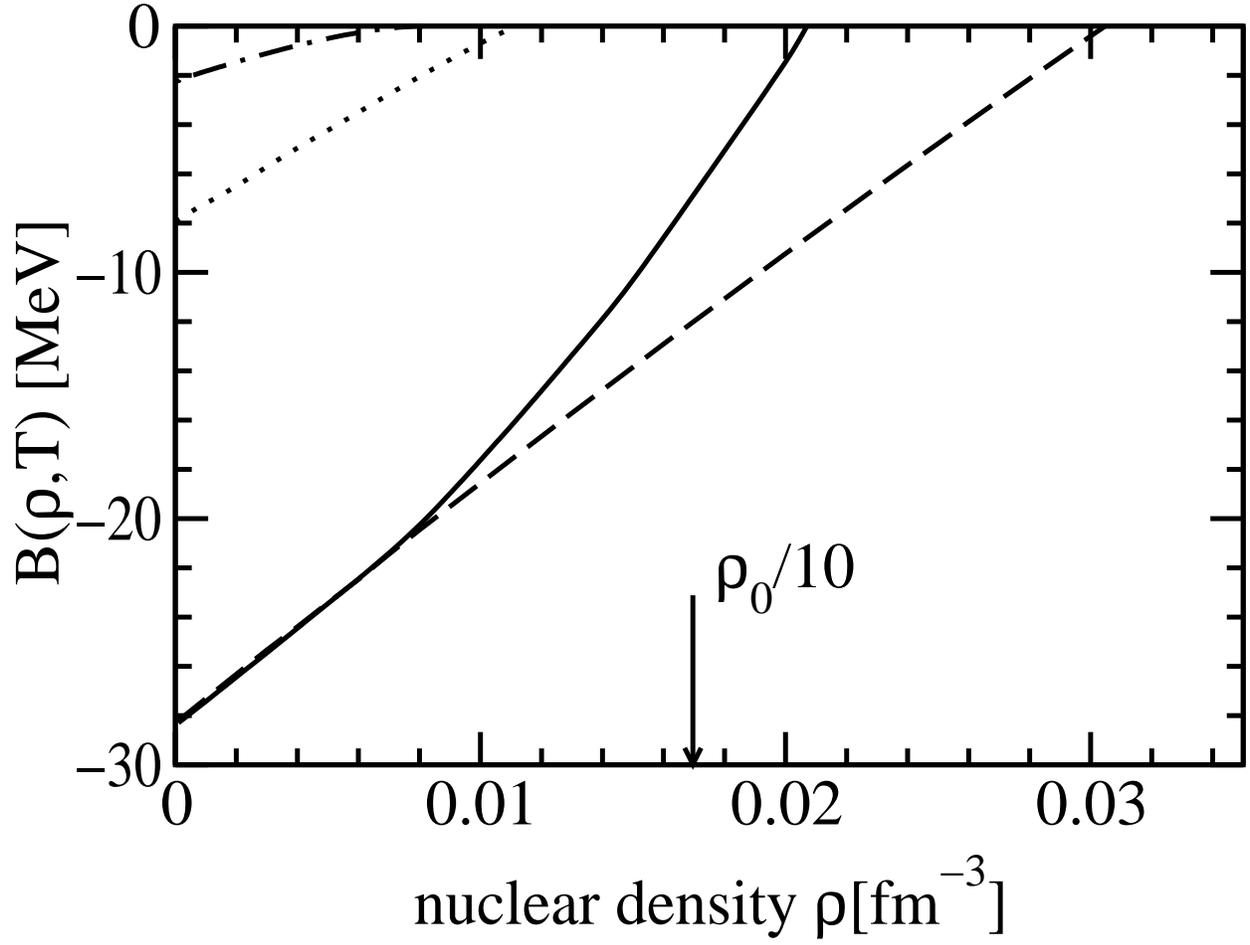}
    \vspace{1cm}
\caption{Binding energy of an $\alpha$-like cluster with zero c.m.
      momentum embedded in symmetric nuclear matter at a temperature
      of $T=10$\,MeV as a function of nucleon density.  Solid line:
      Yamuguchi potential, renormalized to experimental binding energy
      at zero density.  Dashed line: perturbation approach. For
      comparison, the medium dependent binding energies of the
      deuteron (dashed-dotted) and triton (dotted) are also shown.}
    \label{fig:ptop2}
  \end{center}
\end{figure}


\begin{references}
\bibitem{fet71}  
        For a textbook treatment see, e.g., A.L. Fetter, J.D. Walecka,
        {\em Quantum Theory of    Many-Particle Systems},
        (Mc Graw-Hill, New York, 1971).
\bibitem{RMS} 
        G. R\"opke, L. M\"unchow, and H. Schulz,
        Phys. Lett. {\bf B 110}  (1982) 21; Nucl. Phys. {\bf A 379} (1982) 536.
\bibitem{SRS} 
        M. Schmidt, G. R\"opke, and H. Schulz, Ann. Phys. (NY)
          {\bf 202} (1990) 57.
\bibitem{RMSS} 
        G. R\"opke, M. Schmidt, L. M\"unchow, and H. Schulz,
        Nucl. Phys. {\bf A 399} (1983) 587.

\bibitem{bey96} 
        M. Beyer,  G. R\"opke, A. Sedrakian: Phys. Lett. {\bf B 376} (1996) 7.
\bibitem{bey97} 
        M. Beyer and  G. R\"opke,
         Phys. Rev. {\bf C 56} (1997) 2636.
\bibitem{beyFB} 
        M. Beyer, Few Body Syst. Suppl. {\bf 10} (1999) 179.
\bibitem{schadow} 
        M. Beyer, W. Schadow, C. Kuhrts, and G. R\"opke,
        Phys. Rev. {\bf C 60} (1999) 034004.
\bibitem{kuhrts} 
        C. Kuhrts, M. Beyer, and G. R\"opke,  Nucl. Phys. {\bf A} in print.
\bibitem{AGS} 
        E.O. Alt, P.  Grassberger, and W. Sandhas, 
        Nucl. Phys. {\bf B 2} (1967) 167.
\bibitem{duk98} 
        J. Dukelsky, G. R\"opke, and P. Schuck,
        Nucl. Phys. {\bf A 628} (1998) 17.
\bibitem{SchuckR}
        J. Eichler, T. Marumori, and K. Takada, Prog.
  Theor. Phys. {\bf 40} (1968) 60.
\bibitem{SchuckR1} P. Schuck, F. Villars, and P. Ring,
  Nucl. Phys. {\bf A 208} (1973) 302. 
\bibitem{roepke} 
        D. Zubarev, V. Morozov and G. R\"opke,
        {\em Statistical Mechanics of Nonequlibrium Processes I,II},
        (Akademie Verlag Berlin 1996, 1997).
\bibitem{san74} 
        W. Sandhas, Acta Physica Austrica Suppl. {\bf XIII}  (1974) 679.
\bibitem{alt72} 
        E.O. Alt, P. Grassberger, and W. Sandhas, Report
        E4-6688, JINR, Dubna (1972) and in {\em Few-Particle Problems in the
        Nuclear Interaction} eds. I. Slaus et al. (North Holland,
        Amsterdam 1972) p. 299.
\bibitem{san75} 
        W. Sandhas, Czech. J. Phys. {\bf B 25} (1975) 251.
\bibitem{sofianos} 
        S.A. Sofianos, N.J. McGurk, and H. Fiedeldey,
         Nucl. Phys. {\bf A 318} (1978) 295.
\bibitem{alpha} 
        G. R{\"o}pke, A. Schnell, P. Schuck, P. Nozieres,
        Phys. Rev. Lett. {\bf 80} (1998) 3177.
\bibitem{yama} 
        Y. Yamaguchi, Phys. Rev. {\bf 95} (1954) 1628.
\bibitem{gibson}
        B.~F.~Gibson and D.~R.~Lehman,
        Phys.\ Rev.\  {\bf C 18} (1978) 1042.
\end{references}
\end{document}